\documentclass[iop,revtex4]{emulateapj}

\shortauthors{CAVALIERE ET AL.}
\shorttitle{GALAXY CLUSTERS vs. GROUPS}

\slugcomment{ACCEPTED BY APJ}

\begin{document}

\title{The Intragroup versus the Intracluster Medium}
\author{A. Cavaliere$^{1,2}$, R. Fusco-Femiano$^{1}$, and A. Lapi$^{3}$}
\affil{$^1$IAPS-INAF, via Fosso del Cavaliere, 00133 Roma,
Italy - roberto.fuscofemiano@iaps.inaf.it}
\affil{$^2$Accademia dei Lincei, via della Lungara 10, 00165 Roma, Italy.}
\affil{$^3$SISSA, Via Bonomea 265, 34136 Trieste, Italy.}

\begin{abstract}
Galaxy groups differ from clusters primarily by way of their lower masses, $M \sim 10^{14} M_{\odot}$ vs. $M \sim 10^{15} M_{\odot}$. We discuss how mass affects the thermal state of the intracluster or the intragroup medium, specifically as to their entropy levels and radial profiles. We show that entropy is produced in both cases by the continuing inflow of intergalactic gas across the system boundary into the gravitational potential well. The inflow is highly supersonic in clusters, but weakly so in groups. The former condition implies strong accretion shocks with substantial conversion of a large inflow kinetic into thermal energy, whereas the latter condition implies less effective conversion of
lower energies. These features produce a conspicuous difference in entropy deposition at the current boundary. Thereafter, adiabatic compression of the hot gas into the potential well converts such time histories into radial profiles throughout a cluster or a group. In addition, in both cases a location of the system at low $z$ in the accelerating universe or in a poor environment will starve out the inflow and the entropy production,
and produce flattening or even bending down of the outer profile. We analyze in detail the sharp evidence provided by the two groups ESO 3060170 and RXJ1159+5531 that have been recently observed in X rays out to their virial radii, and find a close and detailed match with our expectations.
\end{abstract}


\section{Introduction}

Clusters of galaxies with total masses $M\sim 10^{15} M_{\odot}$ within virial radii $R\sim 2$ Mpc are known (Cavaliere, Gursky \& Tucker 1971, see also Sarazin 1988 for a review of the discovery and of early work) to contain large masses around $M/6$ of a thin, hot proton-electron medium with number densities $n\sim 10^{-3}$ cm$^{-3}$ and temperatures $T\sim 5 - 10 \times 10^7$ K, i.e, particle energies around $kT\sim 5$ keV and total thermal energy up to $10^{65}$ erg.

High $T$ and low $n$ concur to render such an intracluster medium (ICM) an optimal plasma with extreme ratios $kT/e^2 n^{1/3}\sim 10^{12}$ of the average kinetic to the interaction energy, as well as a fluid in local thermal equilibrium with mean free paths $\lambda_{pp}\sim 10\,(kT)^2/n$ and $\lambda_{\rm ep}\sim 40\, \lambda_{pp}$ shorter than the local scales from cluster cores to outskirts. In such conditions thermal, optically thin bremsstrahlung emission produces the intense X rays in the keV spectral region by which the ICM was originally discovered. Meanwhile, the SZ effect (i.e., the Comptonization by the hot ICM electrons of the cold CMB photons crossing the cluster) provides a parallel observational messenger now coming of age (see Cavaliere \& Lapi 2013, henceforth CL13).

The same hot, thin conditions also set in the ICM high values of the specific entropy (actually, the adiabat, see Bower 1997)
\begin{equation}
K \equiv kT/n^{2/3}~.
\end{equation}
This quantity not only records additions to, and radiative losses from the ICM, but also is shown below to play a role in opposing gravity within a potential well (cf. Voit et al. 2005; CL13). Its values are found to range within $K_c\sim 10 - 50$ keV cm$^2$ at the center of a cluster. But outwards of the central $10^2$ kpc, they increase following a radial run close to $K(r)\propto r$ (Tozzi \& Norman 2001, Voit et al. 2005) up to several $10^3$ keV cm$^2$ at the boundary $r = R$ of rich clusters (see Sarazin 1988). Such values far exceed the level $K_0 \sim 10^2$ keV cm$^2$ that prevails in the outer intergalactic medium (IGM) with its density $n_0 \sim 10^{-5}$ cm$^{-3}$ and temperature $T_0$ around $ 10^6$ K (e.g., Branchini et al. 2009), and is conserved on large scales by the adiabatic cosmic expansion.

On the other hand, internal AGNs on account of energetics can materially raise $K$ only in cluster cores or in small groups (see Lapi et al. 2005). So the high internal values throughout the body of clusters and intermediate groups must relate to local gravity. We hold that the gravitational pull of matter already settled in these galaxy systems causes outer intergalactic gas $-$ mixed with outer dark matter (DM) in the cosmic proportion around $0.16$ (see Kravtsov \& Borgani 2012) $-$ to supersonically flow in across the boundary, shock in its vicinity and within $\lambda_{ep}$ thermalize a fair share of its kinetic energy. Thereafter the ICM is adiabatically compressed and stratified in the gravitational potential well into a radial run $K(r)$. The role played by the latter in the final hydrostatic equilibrium of the ICM may be briefly referred to as a modulation of the effective gravitational force, and is detailed in Appendix A (see also Cavaliere et al. 2009, hereafter CLFF09).

A first overall test of this view is provided by recalling that matter
already settled in a cluster sets the virial velocity dispersion at $\sigma \approx (GM /5R)^{1/2} \sim 10^3$ km s$^{-1}$ $\propto M^{1/3}$; this involves also the average proton-electron pairs in the ICM and corresponds to average temperatures $kT \approx \sigma^2 /2 \sim $ keVs as observed. This encourages us to flesh out the above view by appropriately scaling with mass from clusters to groups the conversion from gravitational to thermal energy. Note that already in Cavaliere et al. (1971; see Sarazin 1988) it was proposed that hot, thin conditions ought to hold also in smaller associations of galaxies; X-ray emission from a number of groups was at last established beginning with Kriss et al. (1980) and Schwartz at al. (1980).

Here we take up this lead, and stress that inflows into all such galaxy systems began since the very formation of a protocluster or protogroup at redshifts $z\approx 2-3$ (see Appendix A). Such inflows went on across the once current boundary $R\approx GM/5\sigma^2\propto M^{1/3}$ to accrue about $1/2$ of the present total mass $M$ (see the simulations by Wang et al. 2011). The process implies the present entropy run $K(r)$ to provide a time-resolved view of both the \textit{growth} of the DM binding mass and the heating \textit{history} of the ICM in clusters and groups. Is such a rich view challenged or confirmed by extended and resolved X-ray observations now coming up for galaxy groups with masses $M$ around $M = 10^{14} M_\odot$? This will be the key question tackled in the next sections.

These are organized as follows. In \S~2 we first recap the gravitational production of entropy in clusters and its resulting profile, and then discuss the variants we expect in groups. In \S~3 we analyze the actual conditions observed in groups on using detailed X-ray data from \textit{Suzaku} and \textit{Chandra} concerning two significant instances provided by RXJ1159+5531 (redshift $z$ = 0.081) and ESO 3060170 ($z$ =
0.0358). In \S~4 we discuss our results and draw our conclusions.

Throughout the paper we adopt the standard, flat and accelerating cosmology with round parameters: $H_0 = 70$ km s$^{-1}$ Mpc$^{-1}$, $\Omega_{\Lambda} = 0.7$, $\Omega_M = 0.3$ (Hinshaw et al. 2013; \textsl{Planck} Collaboration XIII 2015). Then 1 arcmin corresponds to $90$ kpc at the distance of RXJ1159+5531, and to $41$ kpc for ESO 3060170.

\section{Gravitational Entropy Production, Clusters vs. Groups}

We first recap the process of gravitational entropy production in the ICM and the resulting entropy profile in rich clusters, to set the stage for dealing next with groups.

\subsection{Clusters}

Toward the cluster boundary at $r= R$, the entropy run takes off from the central value $K_c$ and rises after $K(r)\propto r^a$ with slope given by
\begin{equation}
a = 2.37 - 0.47\, b_R\, ,
\end{equation}
as derived by CLFF09 from matching the gas and the DM inflows so as to retain the cosmic proportion inside. The actual value of $ a$ is modulated by the parameter $b_R \equiv \mu m_p v^2_R/kT_R$. This expresses the ratio at the ICM boundary $R$ of the gravitational to the thermal energy; here $m_p$ is the proton mass and $\mu \approx 0.6$ is the mean molecular weight in the ICM, while $v_R \equiv (GM/R)^{1/2}$ is the DM circular velocity that in the following will normalize the gravitational potential in the form $\phi \equiv \Phi/v^2_R$.

The inflow velocity into a cluster is given in the latter's reference frame by $v = v_R\, (2 \Delta \phi)^{1/2} \sim 10^3 $ km/s in terms of the outer gravitational potential drop $\Delta \phi$ from the turnaround radius $r \approx 2\,R$ where the infall starts to the virial $r \approx R$ (see Appendix A). So the inflow is easily supersonic relative to the sound speed $c_s \equiv (5 kT_0/3 \mu m_p)^{1/2}$ $\sim$ a few $10^2$ km/s in the IGM; correspondingly, the inflow Mach number $\mathcal{M} \equiv v/c_s $ considerably exceeds $1$, so that accretion shocks are set at the boundary.

Specifically, in a layer of thickness $\delta << R$ close to the current virial radius, \textit{strong} shocks will linger with Mach numbers $\mathcal{M}^2 >> 3$, where most of the infall bulk kinetic is converted into random thermal energy (see Appendix B). This yields $kT_R \approx \frac{2}{3} \mu m_p v^2_R \Delta \phi$, in terms of the specific kinetic energy $2\,v^2_R \Delta \phi$ of the gas freely falling across the outer potential drop $\Delta \phi$. Thus we find that $b_R$ may be expressed in two equivalent forms
\begin{equation}
b_R \approx 3 \, v_R^2/c^2_s \mathcal{M}^2 \approx \frac{3}{2\Delta \phi} .
\end{equation}
The first exposes the effect of the inflow Mach number in the regime of strong shocks; it is obtained from using for $kT_R$ the post-shock Rankine-Hugoniot jumps (see Appendix B) that yield $kT_R/kT_o \approx 5\mathcal{M}^2/9$ in the cluster reference frame. The second form exposes the dependence on the outer shape of the gravitational well. It shows that $b_R$ takes on values $2.65 - 2.55$ corresponding to the  parameter $\alpha = 1.27 - 1.3$ governing the Jeans equilibria of the DM, that imply $\Delta \phi \approx 0.57 - 0.59$ (Lapi \& Cavaliere 2009; see also Lokas \& Mamon 2001). Such conditions hold for rich clusters, and there Eqs.~(3) and (2) yield the standard value $a\simeq 1.1$ also obtained by Tozzi \& Norman (2001) from their model.

Here we stress that such steep slopes $a\approx 1$ apply only in
conditions of very effective thermalization of supersonic
infall and entropy production; these occur in strong shocks hovering at the current virial boundary, that allow only small residual bulk kinetic
energy to seep through the shock. Such conditions apply to rich
clusters located in a dense environment that steadily feeds a
smooth inflow, at least in sectors facing dense filaments of the
large scale cosmic network.

On the other hand, all the above conditions may weaken or fail especially in evolved clusters at low $z$ in the accelerating universe; they particularly do in cluster sectors facing a gas-poor region where the inflow has been recently starved. If so, \emph{outer} slope flattening will arise for $r > r_{500}\approx R/2$ (see Appendix A); there $a$ is prone to taper off or even bend down, as indeed is observed in several instances (e.g., Kawaharada et al. 2010; Bonamente et al. 2012; Walker et al. 2012; Ichikawa et al. 2013).

\subsection{Groups}

Here we focus on a different issue, i.e., why in galaxy groups flat slopes  $a < 1$ are to be expected for the entropy run \textit{throughout} the intragroup medium.

In fact, such values $a < 1$ in the body of groups had been hinted at by CLFF09 and CL13, on the basis of parameter values in Eq.~(2) appropriate to groups. In detail, these authors expected $a\approx$ 0.6 - 0.8 to hold throughout the body of groups with $M \lesssim 10^{14} M_\odot$. Next we discuss why such values are to prevail and how they are understood.

The key point is that groups feature generally shallower gravitational potential wells scaled to central depths $\Delta \Phi \sim GM/R \sim M^{2/3}$, and \textit{flatter} runs of the outer potential and of the DM density distributions. Correspondingly, the parameter  governing the Jeans equilibrium (see Appendix A) of the DM takes on values  $\alpha = 1.25$ rather than $1.27 - 1.30$ appropriate for rich clusters; the resulting profiles $\Phi(r)$ are illustrated in Fig.~2 of Cavaliere \& Lapi (2009).

Here we stress that such conditions also imply a decrease of the active
potential drop $\Delta \phi$ over the range $r = 2\,R$ to $R$, so as to drive \textit{lower} Mach number inflows into groups relative to clusters. Not only this feature is relevant by itself, but also it leads to a marked shock outgrowth. This is because weaker inflows reduce the dynamical stresses onto the shock surface, so that the latter outgrows the virial boundary (see Voit et al. 2005) and reduces still more the relevant drop from the turnaround to the shock, to read $\Delta \phi \approx 0.40 - 0.45$ (CLFF09). After Eqs.~(3) and (2) such values imply $a \approx 0.6 - 0.8$.

We emphasize that in groups these conditions and the corresponding \textit{low} values of $a < 1$ held all the way through their formation process, and so left their imprint at all points \textit{throughout} the present structure. In sum, in a group the thermalization efficiency and the related entropy production have always been well below the values appropriate for rich clusters (see Fig.~8 of Cavaliere, Lapi \& Fusco-Femiano 2011), and the correspondingly slopes $a < 1$ remained frozen \textit{everywhere}.

In addition, also a group may inhabit a thin or poor environment due to a very low value of $z$ locating it in the strongly accelerating region of the universe, or due to lack of surrounding large scale filaments. Both conditions concur to cause \emph{additional} flattening out or even bending down of the outer entropy run.

In sum, toward the outer regions of the groups we expect the entropy content to be considerably lower than in clusters.

\section{Evidence}

Remarkably, the rich data set collected by Sun et al. (2009) yielded an average value $<a>\, \sim$ 0.7 beyond 0.15 $r_{500}$, even though with a wide dispersion (see also Bharadwaj et al. 2016, their Fig. 4). Even more remarkably, recent detailed observations concerning ESO 3060170 with  $M \approx 1.7\times 10^{14} M_{\odot}$ have yielded $a\approx 0.81$ (Su et al. 2013), while from those concerning RXJ1159+5531 with its mass $M \approx 0.9\times 10^{14} M_{\odot}$ (see Table 1) we derive $a = 0.66$ as we shall see below.

In fact, a flat slope $a < 1$ results from our analysis of the X-ray data concerning $n(r)$ and $T(r)$ using our detailed SuperModel (SM) . This advanced, entropy-based hydrostatic model, published by CLFF09, incorporates the physical and smooth entropy pattern
\begin{equation}
K(r) = K_c + (K_R - K_c)(r/R)^a
\end{equation}
anticipated in \S~2.1, where $K_c$ and $a$ are treated as two parameters set from fitting the data. A third free parameter is the scale provided by the temperature $T_R$ or the density $n_R$ at the virial radius (see Table 1). We use the virial radii given in the literature (see Table 1); as to the DM `concentration' parameter, we adopt the standard value for groups $c \approx$ 10 (see CL13). Reasonable variations ($\sim 20\%$) of these two fixed parameters have negligible effects on our results. Our model provides the profiles of $n(r)$ and $T(r)$ as detailed in Appendix A. The same hydrostatic model has been used by Su et al. (2015) with many more free parameters in the entropy pattern, and is discussed in \S~4.

Fig.~1 presents our fits to the temperature and density data obtained by Su et al. (2015) from \textit{Chandra} and \textit{Suzaku} X-ray observations of the group RXJ1159+5531 in the North direction. Our entropy profile derived from $K(r) = kT(r)/n(r)^{2/3}$ and shown in Fig.~2 (red line) closely agrees with the entropy data points independently obtained by the above
authors from their observations of $n$ and $T$ in this group. We obtain $a = 0.66\pm 0.03$, well below the value $a = 1.1$ expected for rich clusters, but in line with our group evolution (see \S~2.2). Note that a shallow slope $a<1$ is apparent at an inspection of Fig.~2, and that even shallower slopes are present in other directions as shown by the above authors.

By a similar analysis for ESO 3060170 we obtain $a = 0.87 \pm 0.13$ (see Fig.~3), having tested two different entropy patterns outside the core to describe the X-ray data obtained by the above authors. The first pattern rises in the form of a constant plus a simple power law as described by Eq.~(4); the second (detailed in Appendix C) starts up with an initial slope $a$ as above, and then flattens out for $r> r_b$, where $r_b = (0.57\pm 0.12)R$ is a fitted parameter. The latter pattern closely follows all the entropy points independently obtained by Su et al. (2013), and our value for $a$ is consistent with their $a = 0.81\pm 0.23$. For RXJ1159+5531, on the other hand, out to the outer boundary we found no need or evidence of entropy bending (see blue line of Fig.~2).

\section{Discussion and Conclusions}

We have extended downward to groups with $M \sim 10^{14} M_{\odot}$ our view and our evaluations concerning the thermal state of the ICM in rich galaxy clusters with masses $M\approx 10^{15} M_{\odot}$. Such groups were expected (see CLFF09, CL13) to show considerably \textit{lower} entropy levels throughout, with a run rising across the group body toward the virial boundary at definitely \textit{flatter} slopes than $a = 1.1$. Consistent data have been produced by Sun et al. (2009).

We understand this feature as a sheer effect of the low mass common to all these systems. In clusters the entropy production scales as $\mathcal{M}^2 \propto M^{2/3}$ (see Appendix A) and converges to self-similar conditions, including the entropy slope $a \approx 1.1$. In the group regime the production has to compete with the advected entropy so as to break the self-similar behavior for the total entropy (see Appendix B). Around $M \sim 10^{14} M_{\odot}$ we expect an even flatter radial run; this is just what has been recently observed on comparing the two groups: ESO 3060170 with $M\approx 1.7\times 10^{14} M_{\odot}$ and RXJ1159+5531 with $M \approx 0.9\times 10^{14} M_{\odot}$, see Table 1.

Note that for RXJ1159+5531 our analysis is consistent with Su et al. (2015) for $r >$ 20 kpc (see Fig.~2), i.e., over most of the group. The discrepancy for $r <$ 20 kpc is mainly due to the different density profiles $n(r)$ entering the entropy; it goes back to our fitting their de-projected density data, that yields lower values in the central bins. In fact, the observations of Sun et al. (2009) imply at $r \approx$ 6 kpc an entropy value larger than 15 keV cm$^2$, consistent with our profile.

From Fig.~2 it is easily perceived that the discontinuities in the entropy run as postulated by Su et al. (2015) in the form of a doubly broken power-law implying a total of eleven free parameters are neither really required by the data nor physically expected, as the authors themselves remark in their \S~6.2. They conclude that the outer entropy is consistent with the gravitational value $a$ = 1.1 just as it holds for massive galaxy clusters, based on the intersection at the single point $r_{200}$ of their entropy profile with the straight line $K\propto r^{1.1}$ normalized after Voit et al. (2005).

Out of context, they feel that the properties of the intragroup medium in this fossil group, much less massive and more evolved than any rich cluster, would disfavor the interpretation in terms of recently \emph{starved} accretion shocks proposed by Cavaliere et al. (2011) and reiterated by Fusco-Femiano \& Lapi (2014) in the different context of entropy flattening observed in the outskirts of several massive clusters. At variance with such a claim, in the present context we have shown in \S~2.2 that a flatter slope is to be expected throughout the body of groups as a consequence of weaker inflows always driven by reduced gravity related to their \textit{smaller} masses.

In closer detail, we understand also some outer flattening out of the entropy profile in ESO 3060170, in terms of a recently \emph{starved} inflow to be expected on considering its particularly low $z$ and poor environment noted by Su et al. (2013). Neither circumstance applies to RXJ1159+5531, which in fact shows no evidence of bending down or flattening out. Finally, we understand on the basis of Eqs.~(3) and (2) why the slopes in groups generally show a \textit{wider} intrinsic dispersion (Sun et al. 2009); this is because the values of $b_R$, and even more those of $a$ (given by Eq.~2 as a difference between close terms) are particularly sensitive to relative variations of $\Delta \phi \approx 0.4$, the smaller potential drops that apply to groups. Conversely, the apparent `universality' of $a = 1.1$ in clusters goes back to the limited variance of $b_R$ related to lower relative variance of the larger values $\Delta \phi \approx 0.6$ (see \S~2); this yields a larger and more stable difference between the two terms of Eq.~(2) providing the value $a \simeq 1.1$.

The gravitational origin of shallower entropy slopes within groups is nailed down on comparing the observations of RXJ1159+5531 with those of ESO 3060170, these being the only two fossil groups observed to now with \textit{Suzaku} out to their virial radii. In the latter and more massive group the entropy profile rises in the body with a shallow slope $a \approx 0.8$, but bends down sharply in the vicinity of its virial radius. We stress that such a behavior is just what we \textit{expect} in a system that is twice as massive as RXJ1159+5531, but lies at a lower $z$ and is isolated in a poor environment (see Su et al. 2013). As such, it shares outer conditions with the CC clusters that provide evidence of real outer entropy flattening (see Lapi et al. 2010; Walker et al. 2012; Fusco-Femiano \& Lapi 2014).

If prodded on the specific ground of clusters, we would add that our interpretation of outer flattening of steep slopes in terms of recently \emph{starved} inflow is born out by the azimuthal variations in entropy observed in a number of relaxed clusters (see Ichikawa et al. 2013). Alternative explanations in terms of a clumpy ICM would require quite high values of the clumpiness parameter $C > $ 7 (see Simionescu et al. 2011; Walker et al. 2013; Fusco-Femiano \& Lapi 2013). These are at odds with many updated numerical simulations (e.g., Vazza et al. 2013; Zhuravleva et al. 2013; Roncarelli et al. 2013), and do not show up in probing the conducive outskirts of typical merging clusters like Abell 1750 (Bulbul et al. 2015). On the other hand, the values $C \lesssim 1.6$ at $r_{200}$ as used by Eckert et al. (2015) are far insufficient to explain the entropy bending observed in many clusters since the clumpiness contribution to $K \propto n^{-2/3}$ is just a factor $C^{1/3}\approx 1.2$.

We conclude that on comparing the two groups discussed above, the observed entropy profiles both in the body and at the boundary turn out to behave in detail just as we expect on account of their mass ratio and of their specific redshift/environment differences (see Table 1). In a nutshell, a flat internal slope relates to low $M$, while an outer bending relates to a poor environment.

On the strength of the above multiple evidence, we claim our articulated view concerning the thermal state of the medium filling groups vs. clusters to have been fully confirmed.

\begin{acknowledgements}
We acknowledge useful discussions with F. Gastaldello. We thank our referee for constructive comments that were helpful to sharpen our aim.
\end{acknowledgements}

\appendix

\section{Appendix A}

For the reader's convenience, here we first recall the standard scenario for
the formation of clusters and groups (reviewed in detail by Kravtsov \& Borgani 2012, and used in \S~1). This envisages DM  overdensities initially following the Hubble expansion, then turning around when their size approaches to $2 R$, to collapse and virialize at redshifts $ (1+ z) \propto M^{-2/3}$ or to $M^{-3/4}$ for masses in the cluster or in the group range, respectively.

We add, following Lapi \& Cavaliere (2009) and Wang et al. (2011), that mass inflow continues on and goes mainly into extending the outer halo. The result (used in \S~2 and 3) is a mass density radial profile with a Sersic-Einasto shape. This provides a good analytic approximation to our results for the simplest Jeans equilibria of the DM ($\alpha$ - models, with $\alpha= 1.27 - 1.3$ for clusters and $\alpha = 1.25$ for groups, see Lapi \& Cavaliere 2009), as well as to the outcome of updated numerical simulations (e.g., Wang et al. 2011). These models provide the values of $\Delta\phi$ used in \S~2.

Within such mass distributions, frequently used (as we do in \S~2 and 4) reference sizes comprise an average overdensity $200$ or $500$ higher than the critical density, and read $r_{200} \approx 3\,R/4$ or $r_{500} \approx R/2$, respectively.

The hydrostatic equilibrium of the ICM in the gravitational potential well corresponding to the mass cumulative distribution $M(<r)$ of a cluster or group is expressed by CLFF09 and CL13 in the entropy-based form
\begin{equation}
{{\rm d}p^{2/5}\over {\rm d}r}=-{2\over 5}\,m_p\,G\,{M(<r)\over r^2\, K^{3/5}(r)}\, ,
\end{equation}
where $m_p$ is the proton mass and $G$ the gravitational constant. Eq.~(Al) makes explicit the modulation role played against gravity by the radial entropy run $K(r)$, as anticipated in \S~1.

In terms of $K(r)$ the $3D$ density $n(r)$ and the temperature $T(r)$ follow, since $K\propto T/n^{2/3}$ (see Eq.~1) and $p \propto n\, T$ hold. These observables may be fitted on the data, to fix any parameters entering the pattern assumed for $K(r)$. In addition, the electron pressure may be directly obtained from the SZ effect, see CL13 and references therein. Such a non linear rise yields the threshold $M\approx 10^{14}\, M_\odot$ stated in \S~4.

\section{Appendix B}

Next, to complement \S~2 we write the Rankine-Hugoniot jumps in the cluster/group reference frame for the post-shock relative to the pre-shock
temperature and entropy. These are obtained from conservation of mass, momentum and energy across the shock surface at $r \sim R$, in fact, over the thickness of an electron mean free path $\lambda_{ep} << R$ (where $\lambda_{ep} = \sqrt{m_p/m_e}\times \lambda_{pp}$, see Zel'dovich \& Raizer 1967) complete equilibrium is achieved.

Then the jumps (used in \S~2 and 3) are given by CLFF09 (see Eqs.~B4 and B5) in the form $kT_R/kT_0 \approx 5\mathcal{M}^2/9 + 3/2 + \mathcal{O}(\mathcal{M}^{-2})$ for the temperature. As for entropy, the jump approaches $K_R/K_0 \approx 5\mathcal{M}^2/(9\times 4^{2/3}) + 17/(8\times 4^{2/3})\approx 0.22\mathcal{M}^2+0.84$, with the last term giving the contribution simply advected by the IGM inflow (see also Voit et al. 2003); net entropy production requires $\mathcal{M}^2\ga 3.8$, and takes off steeply for $\mathcal{M}^2\gg 3$, in keeping with a classic result in the physics of fluids (Landau \& Lifshitz 1959, see also Fig.~3 of CL13). In such conditions relevant for clusters, Eq.~(2) yields $a\approx 1$ through most of the body.

On the other hand, the slope provided by Eq.~(2) drops rapidly to $a\approx 0.6$ and below in moving to groups with $M\la 10^{14}\, M_\odot$ where $\Delta\phi\la 0.4$ holds. Then central AGN heating or radiative cooling may intermittently prevail, and enhance the variance in slope values.

Realistically, in all cases the transition from the outer IGM to the inner ICM will take place at the boundary in a patchwork of shocks within a layer of thickness $\delta << R$. From conservation of mass, momentum and energy across the layer, the jumps will now contain additional terms $\mathcal{O}(\delta/R)$ that may be taken to also include effects of non-spherical geometry (Lapi, Cavaliere, \& Menci 2005). For $\delta/R << 1$ the classic Rankine-Hugoniot forms are recovered.

\section{Appendix C}

For the SM analysis of the group ESO 3060170 discussed in \S~3 the value of $a$ applies in the body for $r \leq r_b$, but for $r > r_b$ it declines following $a - s(r/r_b -1)$ with a costant gradient $s$. The full entropy profile then reads
\begin{equation}
k(r)=\left\{
\begin{array}{lll}
K_c + (K_b - K_c)(r/r_b) & \; \; r\leq r_b \\
\\
K_R(r/R)^{a+s} e^{s(R - r)/r_b} & \; \; r > r_b
\end{array}
\right. .
\end{equation}
where $K_b = r_b^{(a+s)} e^{s(1.0-r_b)/r_b}$. The outer branch describes a simple, linear decline of the slope with the gradient $s$; normalizations have been set so as to obtain a continuous function and derivative for $K(r)$ (see Lapi, Fusco-Femiano, \& Cavaliere 2010).

\clearpage
\begin{figure}
\centering
\epsscale{0.5}\plotone{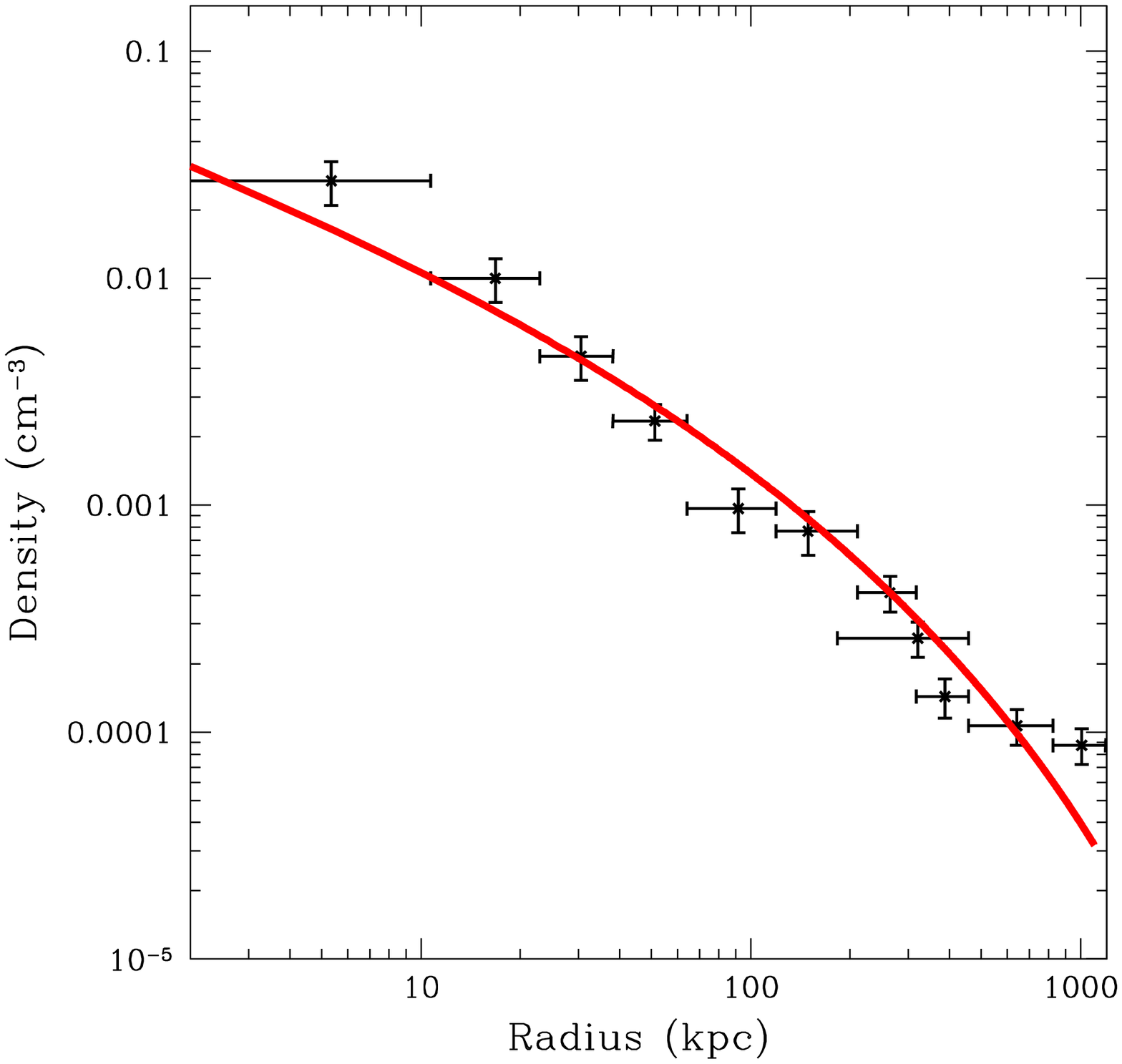}\\\plotone{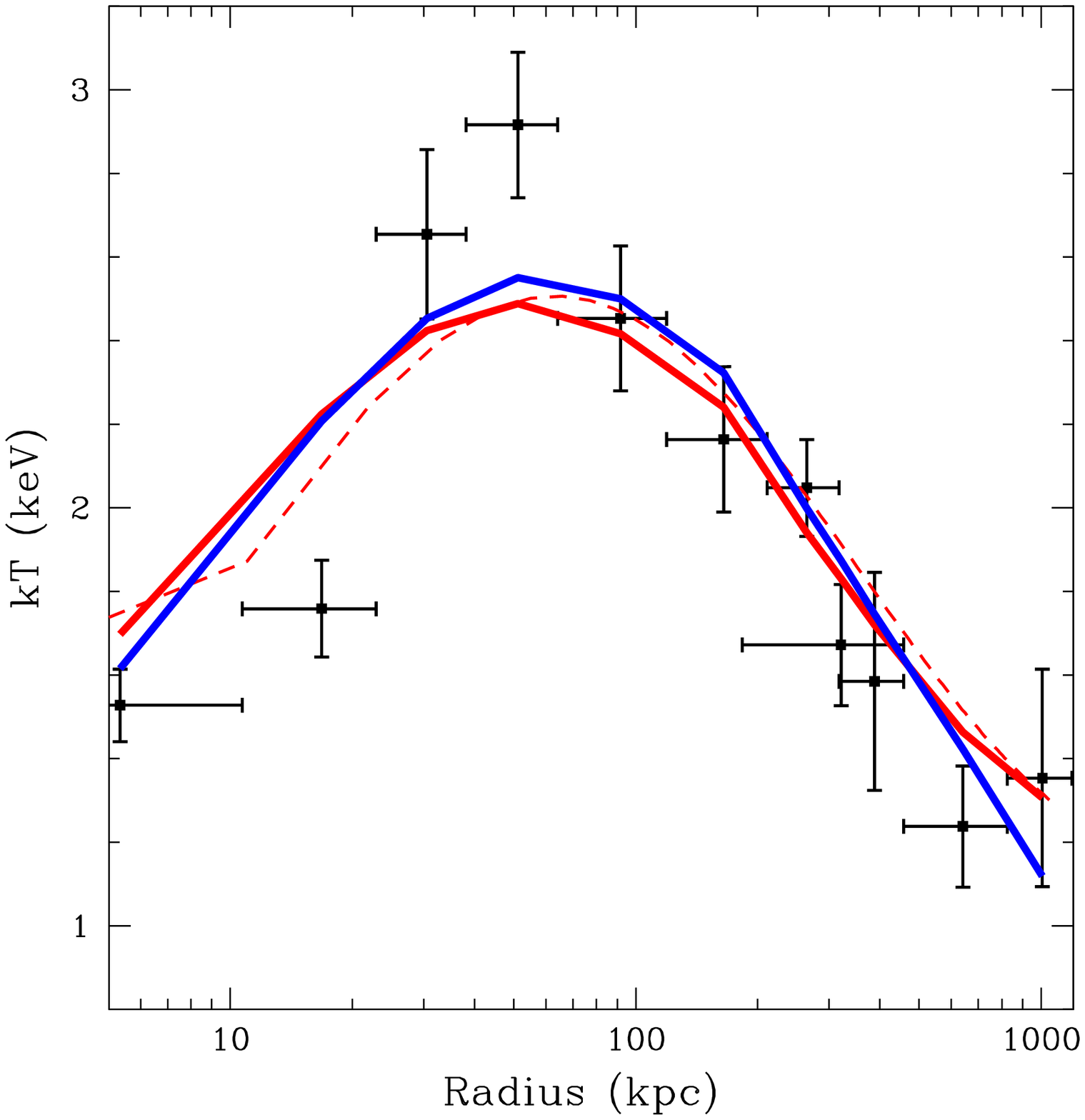}
\caption{Top panel: 3D density data in the North direction of the `fossil' group RXJ1159+5531, obtained via the spectral deprojection tool \textsf{projct} by Su et al. (2015). The red line presents our SuperModel (CLFF09) fit to the data obtained on using the simple, physical entropy pattern of Eq.~(4), with $R$ given in Table 1; the reduced $\chi^2$ is 11.1/8 = 1.4. Bottom panel: Projected temperature profiles of RXJ1159+5531 measured by the above authors in the North direction with \textit{Chandra} and \textit{Suzaku}; the red line presents our SM fit to this data using the above entropy pattern with the reduced $\chi^2= 12.9/8 = 1.6$; the blue line is our SM fit, adding an entropy flattening for $r > r_b$ (as detailed in Appendix C). The red dashed line shows the de-projected temperature profile corresponding to the red line.}
\end{figure}

\clearpage
\begin{figure}
\centering
\epsscale{0.7}\plotone{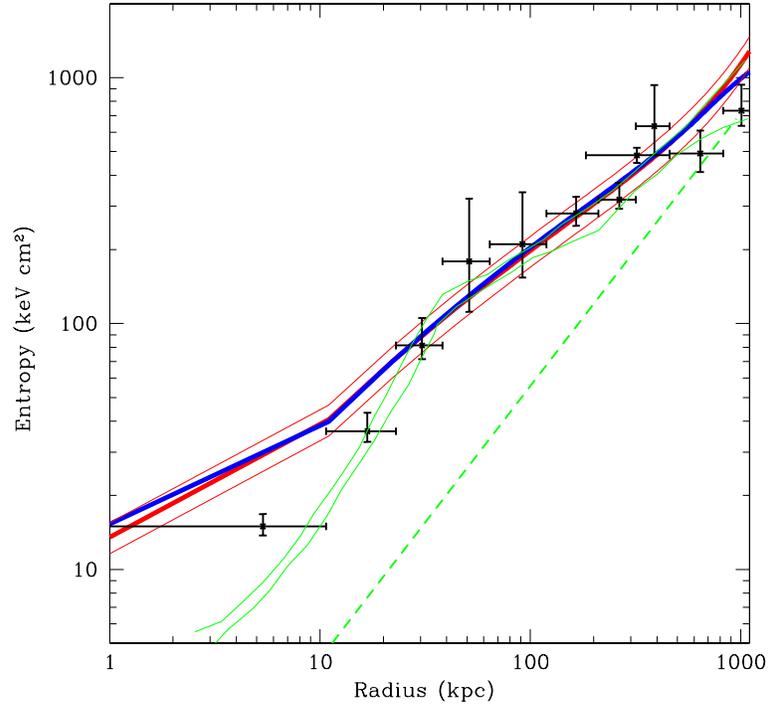}\caption{The red line shows the entropy profile we derive for RXJ1159+5531 from the de-projected temperature (red dashed line of Fig.~1) and density profiles obtained with our SM analysis of the X-ray data given in Fig.~1; the thin red lines show the 1$\sigma$ uncertainties. The blue line is obtained from the de-projected temperature profile given by the blue line of Fig.~1, and shows the absence of any entropy bending in such a group. The red profile closely complies with the entropy points independently derived from the X-ray data by Su et al. (2015). For comparison, the dashed green line shows the profile with uniform slope $a$ = 1.1 normalized at its upper end as in Su et al. (2015). The solid green lines report the profile with its $1\sigma$ uncertainties as obtained by the above authors on using their analytical run for \textit{K(r)} broken into three segments with different slopes.}
\end{figure}

\clearpage
\begin{figure}
\centering
\epsscale{0.7}\plotone{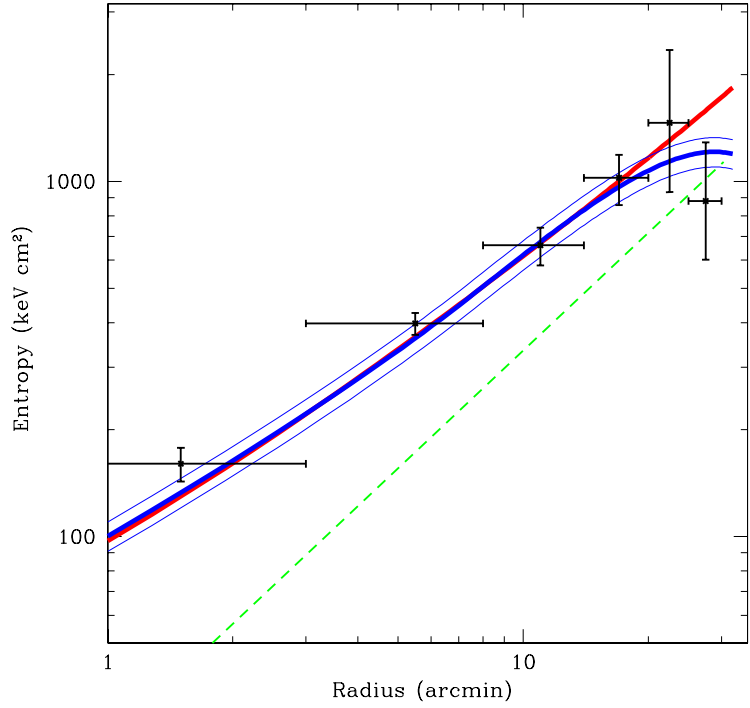}\caption{For the `fossil'
group ESO 3060170 the blue and red lines show the entropy profile
we derive from the temperature and density profiles obtained with
our SM analysis of the X-ray data observed by Su et al.
(2013); the thin blue lines show the 1$\sigma$ uncertainties.
The red line is obtained on assuming the pattern in Eq.~(4),
while for the blue line we assume a pattern that again starts up with slope $a$, but then bends down for $r > r_b$ (details in Appendix C).
The blue entropy profile closely complies with all points
independently derived from the X-ray data. For comparison, the
dashed green line has a slope $a$ = 1.1 and is normalized as in Fig. 2. We adopted $R$ = 32$^{\prime}$ following Sun et al. (2004).}
\end{figure}

\clearpage
\centerline{Table 1: Main features and results concerning two fossil Groups}
\begin{center}
\begin{tabular}{ccc}
\hline
Name   & ESO 3060170 & RXJ1159+5531 \\
\hline
            &   &  \\
$M/10^{14} M_{\odot}$  & 1.7$^{(1)}$ & 0.9$^{(2)}$ \\
\medskip
$R$ (kpc) & 1312$^{(3)}$ & 1100$^{(2)}$ \\
\medskip
$z$ & 0.036$^{(1)}$ & 0.081$^{(2)}$ \\
\medskip
environment & poor$^{(1)}$  & average$^{(2)}$ \\
\medskip
$a$ & $0.87\pm 0.13$ & $0.66\pm 0.03$ \\
\medskip
$K_c$ (keV cm$^2$) & 23$\pm$3 & $<$ 0.8 \\
\medskip
$kT_R$ (keV) & 0.96 $\pm$ 0.08 & 1.29 $\pm$ 0.20 \\
\medskip
$n_R$ (cm$^{-3}$) & (2.27$\pm$0.04)$\times 10^{-5}$ & (3.19$\pm$0.05)$\times 10^{-5}$ \\
\medskip
$r_b$ (kpc) & (0.57 $\pm$ 0.12) $R$ & --- \\
\medskip
$s$ & 0.49 $\pm$ 0.10 & --- \\
\hline
\end{tabular}
\end{center}
\par\noindent
\centerline{$^{(1)}$ Su et al. (2013); $^{(2)}$ Su et al. (2015); $^{(3)}$ Sun et al. (2009). All other values are obtained in the present work.}

\end{document}